\documentclass [11pt] {rspublic}
\usepackage{amsfonts,latexsym,amsmath,amssymb,amsthm}
\usepackage[usenames]{color}
\usepackage{graphicx}
\graphicspath{{figs/}}
\usepackage{lineno}
\usepackage{natbib}

\begin{document}
%\setcitestyle{notesep={; },round,aysep={},yysep={,}}
\bibliographystyle{plainnat}

\def\bzero{\mathbf{0}}
\def\bone{\mathbf{1}}
\def\bigO{\mathcal{O}}
\def \diag{\mbox{diag}}
\def \I{{\mathcal I}}
\def \W{W}
\def \x{\mathbf{x}}
\def \N{\mathbf{N}}
\def \R{\mathbf{R}}
\def\Rp{\mathbb{R}_+}
\def \A{\mathbf{A}}
\def \D{\mathbf{D}}
\def \B{\mathbf{C}}
\def \v{\mathbf{v}}
\def \a{\mathbf{a}}
\def \b{\mathbf{b}}
\def \c{\mathbf{c}}
\def \r{\mathbf{r}}
\def \m{\mathbf{m}}

\title[Coexistence of intransitive metacommunities]{Spatial heterogeneity promotes coexistence of rock-paper-scissor metacommunities}

\author[S.J. Schreiber \& T.P. Killingback]{Sebastian J. Schreiber$^1$ and Timothy P. Killingback$^2$}
\affiliation{$^1$Department of Evolution and Ecology and the Center for Population Biology
University of California, Davis, California 95616 USA, e-mail: sschreiber@ucdavis.edu\\
$^2$Department of Mathematics, University of Massachusetts, Boston, Massachusetts 02125 USA, e-mail: timothy.killingback@umb.edu}

\maketitle
\begin{abstract}{rock-scissors-paper game, non-transitive interactions, evolutionary game
theory, replicator dynamics, spatial heterogeneity, metacommunity dynamics}
\textbf{Abstract.} \baselineskip 15pt 
The rock-paper-scissor game -- which is characterized by three strategies R,P,S, satisfying the non-transitive relations S excludes P, P excludes R, and R excludes S -- serves as a simple prototype for studying more complex non-transitive systems. For well-mixed systems where interactions result in fitness reductions of the losers exceeding fitness gains of the winners, classical theory predicts that two strategies go extinct. The effects of spatial heterogeneity and dispersal rates on this outcome are analyzed using a general framework for evolutionary games in patchy landscapes. The analysis reveals that  coexistence is determined by the rates at which dominant strategies invade a landscape occupied by the subordinate strategy (e.g. rock invades a landscape occupied by scissors) and the rates at which subordinate strategies get excluded in a landscape occupied by the dominant strategy (e.g. scissor gets excluded in a landscape occupied by rock). These invasion and exclusion rates correspond to eigenvalues of the linearized dynamics near single strategy equilibria. Coexistence occurs when the product of the invasion rates exceeds the product of the exclusion rates.  Provided there is sufficient spatial variation in payoffs, the analysis identifies a critical dispersal rate $d^*$ required for regional persistence. For dispersal rates below $d^*$, the product of the invasion rates exceed the product of the exclusion rates and the rock-paper-scissor metacommunities persist regionally despite being extinction prone locally. For dispersal rates above $d^*$, the product of the exclusion rates exceed the product of the invasion rates and the strategies are extinction prone. These results highlight the delicate interplay between spatial heterogeneity and dispersal in mediating long-term outcomes for evolutionary games.

\textbf{Author's Summary:}  \baselineskip 15pt  The rock-paper-scissor game, which might initially seem to be of purely theoretical interest, plays an important role in describing the behavior of various real-world systems including the evolution of alternative male mating strategies in the side-blotched lizard, the evolution of bacterial populations, and coexistence in plant communities. While the importance of dispersal in mediating coexistence for these intransitive communities has been documented in theoretical and empirical studies, these studies have, by in large, ignored the role of spatial heterogeneity in mediating coexistence. We introduce and provide a detailed analysis of models for evolutionary games in a patchy environment. Our analysis reveals that spatial heterogeneity coupled with low dispersal rates can mediate regional coexistence, despite species being extinction prone in all patches. The results suggests that diversity is maintained by a delicate interplay between dispersal rates and spatial heterogeneity.  
\end{abstract}
\section*{Introduction}

\linenumbers
\baselineskip 24pt

Since its inception over 30 years ago evolutionary game theory has become a major theoretical framework for studying the evolution of frequency dependent systems in biology \citep{maynardsmith-82,hofbauer-sigmund-98,hofbauer-sigmund-03}. There have been numerous applications of evolutionary game theory in biology (and increasingly also in economics and the social sciences), ranging from the evolution of cooperation \citep{axelrod-84,axelrod-hamilton-81} and animal conflicts \citep{maynardsmith-price-73}, to the evolution of sex ratios \citep{hamilton-67}, and the origin of anisogamy \citep{parker-etal-72}. Indeed it is striking that three of the simplest possible games that can be considered, the Prisoner's Dilemma game \citep{axelrod-84}, the Hawk-Dove (or Snowdrift) game \citep{maynardsmith-82}, and the Rock-Paper-Scissor game \citep{hofbauer-sigmund-98}, have all found fruitful applications in the study of important biological problems, namely, the evolution of cooperation \citep{axelrod-84,axelrod-hamilton-81}, the evolution of animal contests \citep{maynardsmith-82,maynardsmith-price-73}, and the evolution of Red Queen  dynamics \citep{sinervo-lively-96,kerr-etal-02,kirkup-riley-04} (in which the system cycles constantly between the different possible strategies).
 
In formulating evolutionary game theory it is often assumed that the individual strategists interact at random in a well-mixed population. Under this assumption the evolutionary game dynamics can be formulated as a system of ordinary differential equations, the replicator equations, which describe the time evolution of the different strategies in the game \citep{maynardsmith-82,hofbauer-sigmund-98}. Any evolutionary stable strategies (i.e. a strategy, which if adopted by almost all members of the population, cannot be invaded by any mutant strategy) is a stable equilibrium of the replicator equations \citep{hofbauer-sigmund-98}.  

In many situations the assumption that the population is well-mixed, with individuals interacting randomly throughout the whole population, is not realistic. This will often be the case if there is some spatial structure in the population, which results in individuals interacting more with neighboring individuals than with more distant ones. One way of modeling a structured population is to assume that individuals are associated with the vertices of a graph, with two individuals interacting if the corresponding vertices are connected by an edge. This approach leads to a network based formulation of evolutionary game theory in which the evolutionary dynamics on the graph is determined by a suitable deterministic or stochastic analogue of the replicator dynamics. Evolutionary games on graphs have been rather well studied \citep{nowak-may-92,killingback-doebeli-96,nakamaru-etal-97,hauert-szabo-03,ifti-etal-04,hauert-doebeli-04,santos-pacheco-05,ohtsuki-etal-06}. One of the basic conclusions of this work is that the evolutionary dynamics of a game on a graph can be quite different from the dynamics of the game in a well-mixed population. A particularly important instance of this is that cooperation can be maintained in the Prisoner's Dilemma game on a graph. In contrast,  in a well-mixed population cooperation is always driven to extinction by defection.
 
 An alternative way of modeling a structured population is to assume that it is composed of a number of local populations, within which individuals interact randomly, coupled by dispersal. In this approach the total population or community is modeled as a metapopulation or metacommunity. Metapopulation and metacommunity structure is known to have important implications for population dynamics in ecology and evolution \citep{hanski-99,holyoak-etal-05,prsb-10}.
 
In spite of the considerable amount of work that has been devoted to understanding the ecological and genetic consequences of metacommunity structure there has been much less attention devoted to studying the dynamics of evolutionary game theory in the metacommunity context. The purpose of this paper is to provide a general mathematical formulation of metacommunity evolutionary game dynamics, and to obtain detailed results for the case of a particularly interesting game -- the rock-paper-scissors game. In the last few years the rock-paper-scissor game, which might initially seem to be of purely theoretical interest, has emerged as playing an important role in describing the behavior of various real-world systems. These include the evolution of alternative male mating strategies in the side-blotched lizard \emph{Uta Stansburiana} \citep{sinervo-lively-96}, the \emph{in vitro} evolution of bacterial populations \citep{kerr-etal-02,nahum-etal-11}, the \emph{in vivo} evolution of bacterial populations in mice \citep{kirkup-riley-04}, and the competition between genotypes and species in plant communities~\citep{lankau-strauss-07,cameron-etal-09}. More generally, the rock-scissors-paper game -- which is characterized by three strategies R, P and S, which satisfy the non-transitive relations: P beats R (in the absence of S), S beats P (in the absence of R), and R beats S (in the absence of P) -- serves as a simple prototype for studying the dynamics of more complicated non-transitive systems  \citep{buss-jackson-79,paquin-adams-83,may-leonard-75,jmb-97,oikos-04,vandermeer-pascual-05,allesina-levine-11}.

One of the central issues that has arisen in recent years in ecology is the degree to which metacommunity structure can lead to the coexistence of competing species \citep{hanski-99,amarasekare-nisbet-01,moquet-etal-05, gravel-etal-10}. Here, we study an interesting aspect of this larger question, namely, the effect of a general metacommunity structure on the coexistence of the strategies in the rock-paper-scissor game. In a well-mixed population the evolutionary dynamics of the rock-paper-scissor game is known to be determined by the sign of the determinant of the payoff matrix \citep{hofbauer-sigmund-98}. If the determinant of the payoff matrix is positive then the replicator dynamics converges to a stable limit point, in which the frequencies of the three strategies tend to constant values. If, however, the determinant of the payoff matrix is negative then the replicator dynamics converges to a heteroclinic cycle, in which the frequencies of the three strategies continue to undergo increasingly extreme oscillations. In the latter case the frequencies of the different strategies successively fall to lower and lower levels as the population dynamics approach the heteroclinic attractor. Consequently,   stochasticity would result in the ultimate extinction of one of the strategies followed by the elimination of the remaining dominated strategy.

In this paper we study the dynamics of the rock-scissors-paper game in a metacommunity context, and show that dispersal in spatially heterogeneous environments can alter dynamical outcomes. In particular, we characterize under what conditions dispersal in heterogeneous environments stabilizes or destabilizes rock-paper-scissor metacommunities. When dispersal is stabilizing, all strategies in the  rock-scissors-paper metacommunity are maintained indefinitely by a Red Queen type dynamic.

\section*{Model and Methods}
\setcounter{equation}{0}
\renewcommand{\theequation}{\arabic{equation}}
\subsection*{Evolutionary Games in Space.}
We consider interacting populations playing $m$ distinct strategies ($i=1,\dots,m$) in a spatially heterogeneous environment consisting of $n$ patches ($r=1,\dots,n$). Space is the primary limiting resource for the populations and assumed to be fully saturated i.e. all sites within a patch are occupied. Let $x_i^r$ denote the frequency of strategy $i$ in patch $r$. Within patch reproductive rates of individuals are determined by pairwise interactions where an individual in patch $r$ playing strategy $i$ receives a payoff of  $A_{ij}(r)$  following an encounter with an individual playing strategy $j$. Individuals reproduce at a rate equal to their net payoff. For individuals playing strategy $i$ in patch $r$, this net payoff equals $\sum_i A_{ij} (r) x_j^r$. All individuals in patch $r$ experience a per-capita mortality rate $m^r$. Dying individuals free up space that can be colonized with equal likelihood by all offspring living in the patch. In the absence of dispersal, the probability that a site emptied by a dying individual gets colonized by an offspring playing strategy $i$ is $\frac{\sum_i A_{ij} (r) x_j^rx_i^r}{\sum_{j,k} A_{jk}(r) x_j^r x_k^r}$. Thus, in the absence of dispersal, the  population dynamics in patch $r$ are 
\begin{equation}\label{local}
\frac{dx^r_i}{dt} = -m^r\,x_i^r + m^r \frac{\sum_j A_{ij}(r) x_i^r x_j^r}{\sum_{j,k} A_{jk}(r) x_j^r x_k^r}.
\end{equation}

To account for movement between patches, let $d_{sr}$ denote the fraction of progeny born in patch $s$ that move to patch $r$. In which case, the rate at which offspring of strategy $i$ arrive in patch $r$ equals $\sum_s d_{sr} \sum_j A_{ij}(s) x_i^s x_j^s$ and the probability an offspring playing strategy $i$ colonizes an emptied site equals $ \frac{\sum_j A_{ij}(s) x_i^s x_j^s}{\sum_s d_{sr}\sum_{j,k} A_{jk}(s) x_j^s x_k^s}$. Hence, the full spatial dynamics are
\begin{equation}\label{replicator}
\frac{dx^r_i}{dt} = -m^r\,x_i^r + m^r \frac{\sum_s d_{sr} \sum_j A_{ij}(s) x_i^s x_j^s}{\sum_s d_{sr}\sum_{j,k} A_{jk}(s) x_j^s x_k^s}.
\end{equation}
We assume that the matrix $D$ of dispersal probabilities is primitive
(i.e. after sufficiently many generations, the decedents of any individual in any one patch  occupy all patches).

For the rock-paper-scissor game, there are three strategies with rock as strategy $1$, paper as strategy $2$, and scissor as strategy
$3$. Let $a^r$ be the basal reproductive rate of an individual in patch $r$. Let
$b_i^r$ (i.e. the benefit to the winner)  be the payoff to
strategy $i$ in patch $r$ when it wins against  its subordinate
 strategy, and $-c_{i}^r$ (i.e. the cost to the loser) be the payoff to strategy $i$
 in patch $r$ when it loses against the dominant strategy. Under these
 assumptions, the payoff matrix in patch $r$ is given by
\begin{equation}\label{payoff}
\A(r)=a^r+\begin{pmatrix}
0& -c_1^r& b_1^r \\
b_2^r& 0 & -c_2^r \\
-c_3^r & b_3^r & 0
\end{pmatrix}
\end{equation}
Throughout this article, we assume that $a^r>0$, $0<c_i^r<a^r$,
$b_i^r>0$.  The assumption $a^r>c_i^r$ ensures that payoffs remain positive. 

\subsection*{Analytical and Numerical Methods} To understand whether the strategies persist in the long-term,
we analyze (\ref{replicator}) using a combination
of analytical and numerical methods. Long-term persistence of all the strategies
is equated with \emph{permanence}: there exists a
minimal frequency $\rho>0$ such that
\[
x_i^r(t) \ge \rho \mbox{ for all $i,r$}
\]  
whenever $t$ is sufficiently large and all strategies are initially present (i.e. $\sum_r x_i^r (0)>0$ for
$i=1,2,3$).  Permanence ensures that populations  recover from rare large perturbations and
continual small stochastic perturbations \citep{dcds-07,benaim-etal-08}. Using analytical techniques developed by \citet{jde-10}, we derive an analytical condition for permanence in terms of products of eigenvalues at the single strategy equilibria of the model. These criteria take on an explicit, interpretable form  when (i) populations are relatively sedentary (i.e. $d_{rr} \approx 1$ for all $r$) and (ii) populations are well mixed (i.e. there exists a probability vector $v=(v_1,\dots,v_n)$ such that $d_{rs}\approx v_s$ for all $r,s$).    To better understand permanence at intermediate dispersal rates, we derive an analytical result about critical dispersal thresholds for persistence of metacommunity exhibiting unconditional dispersal (i.e probability of leaving a patch is independent of location) and numerically simulate (\ref{replicator}) using the deSolve package of R~\citep{r}. To simplify our exposition, we present our results under the assumption that $m^r=m$ and $a^r=a$ for all $r$ i.e. there is only spatial heterogeneity in the benefits and in the costs. More general results are presented in the Appendices. 

\section*{Results}
 
\setcounter{equation}{3}
\renewcommand{\theequation}{\arabic{equation}}

\begin{figure}[t]
\begin{center}
\begin{tabular}{cc} 
\includegraphics[width=2in]{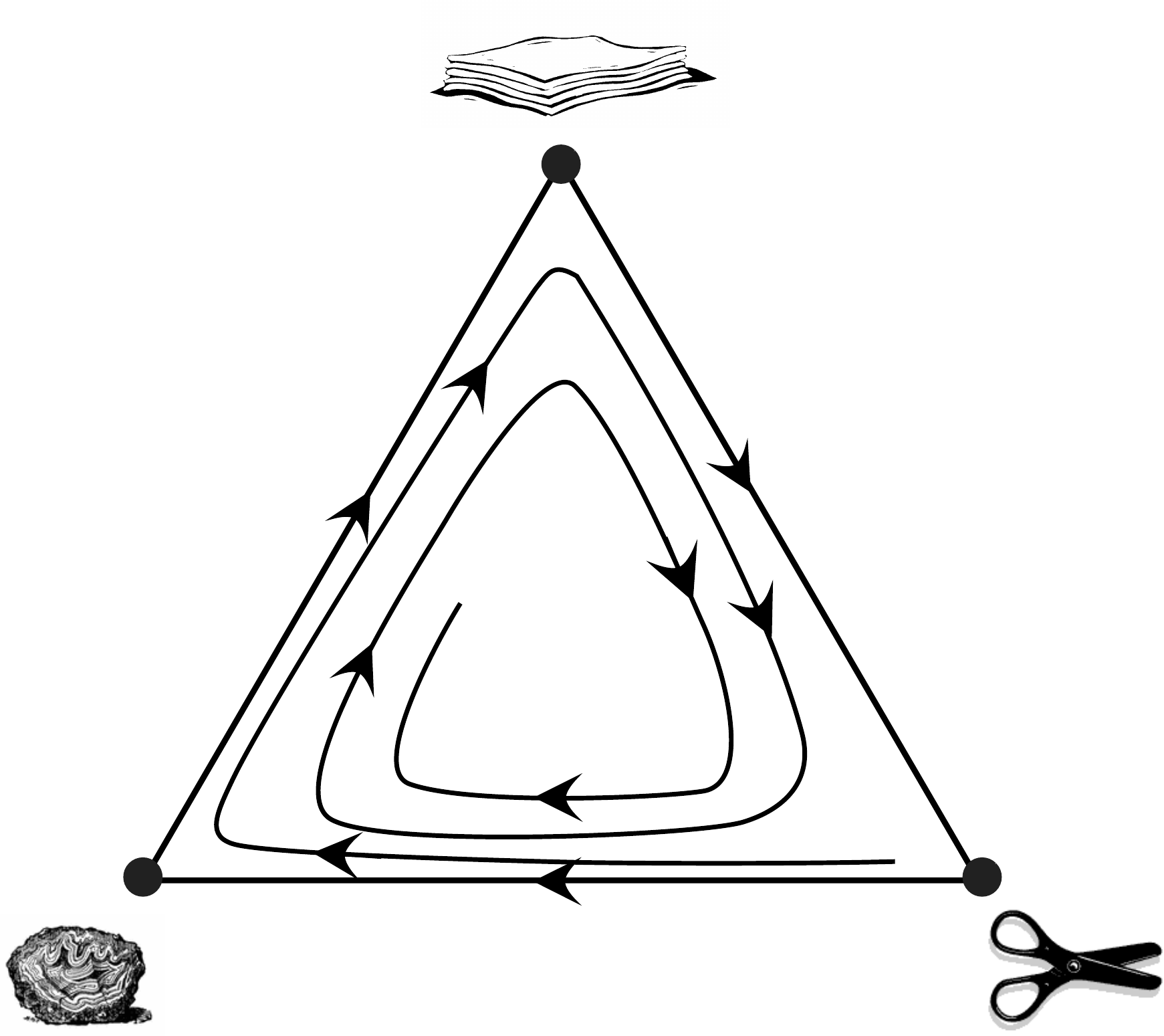}& \includegraphics[width=2in]{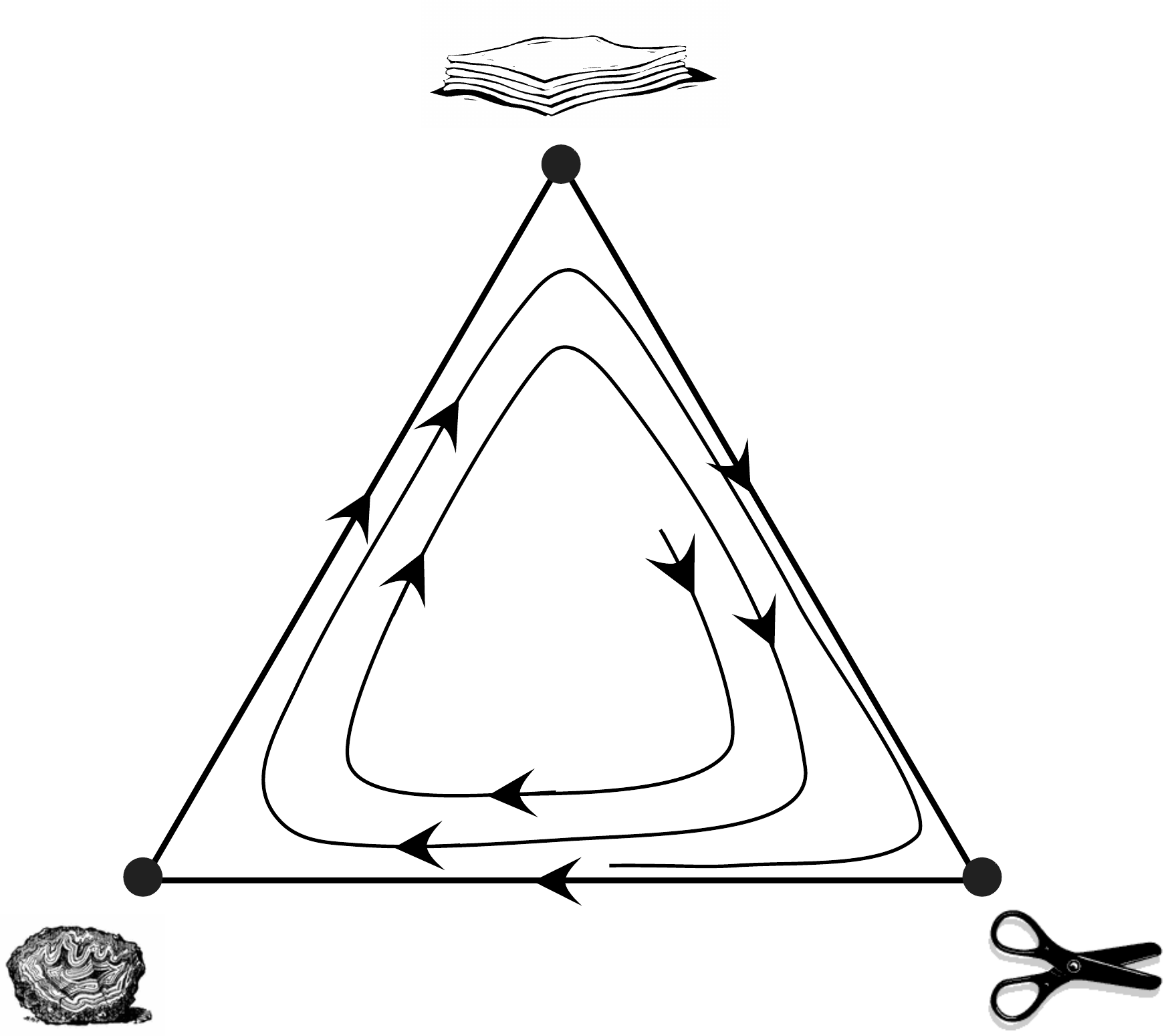}\\ 
\includegraphics[width=2.75in,trim=0 0 0 1in, clip]{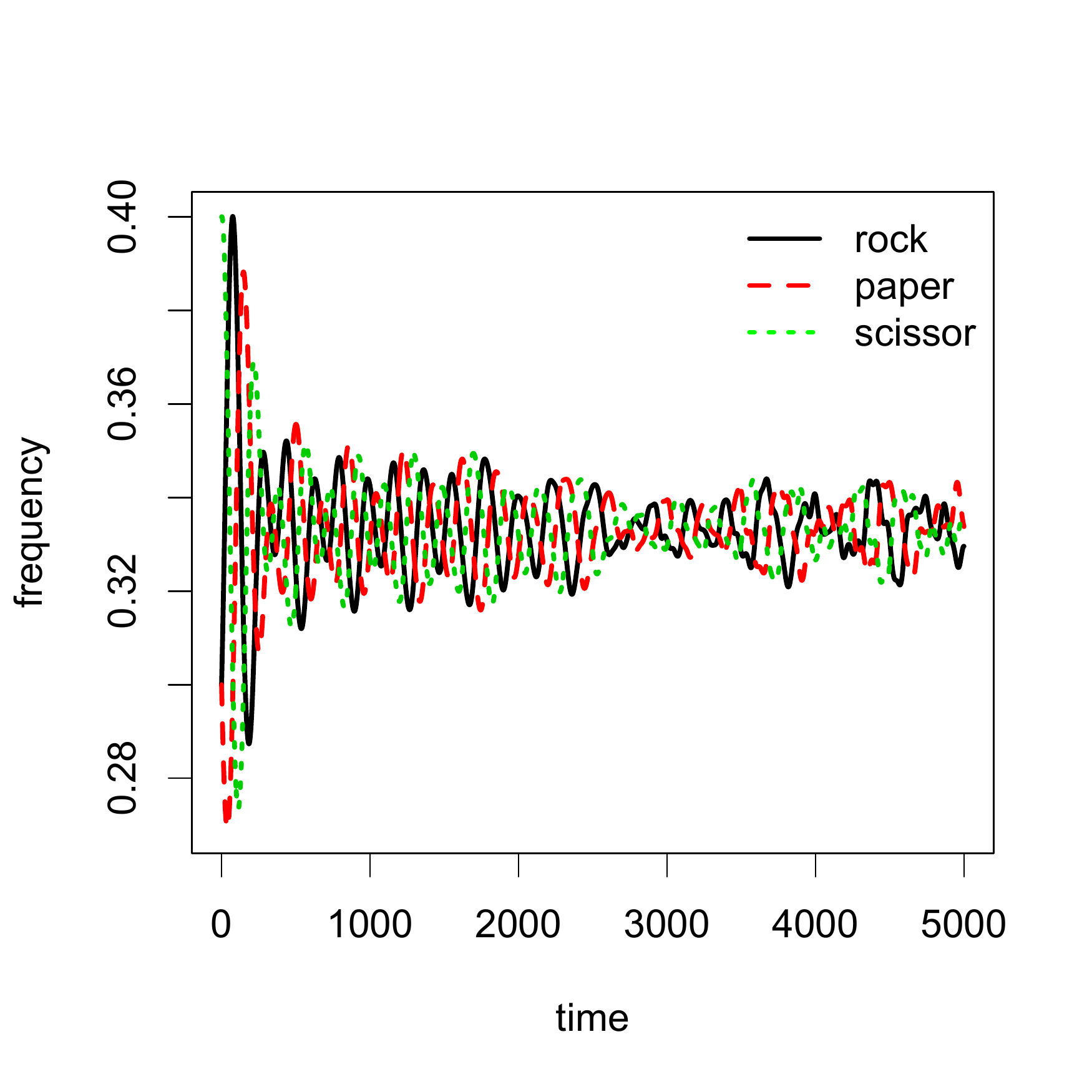}& \includegraphics[width=2.75in,trim=0 0 0 1in, clip]{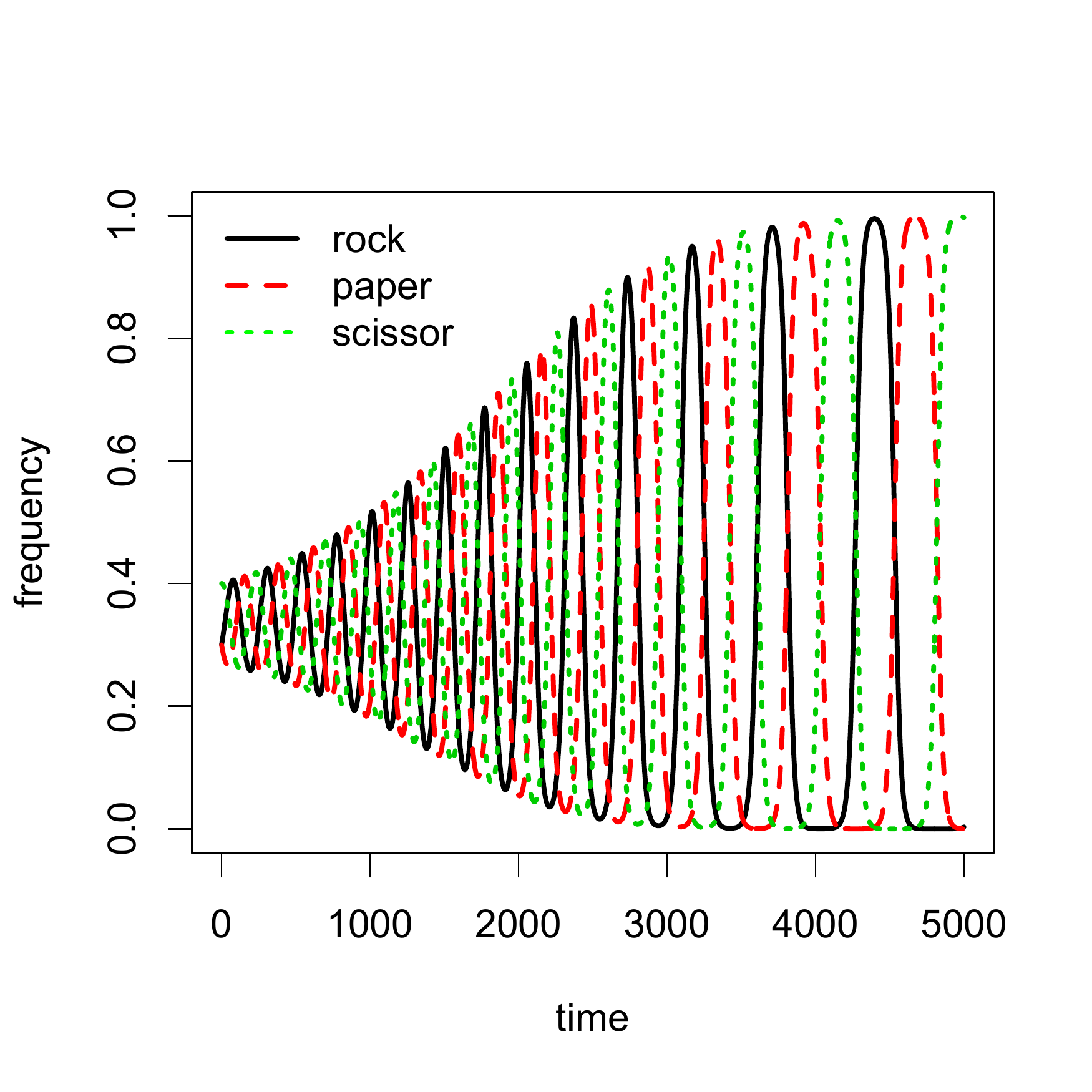}\\
(a) repelling cycle& (b) attracting cycle
\end{tabular}
\caption{Boundary dynamics for rock-paper-scissors. For within patch and metacommunity dynamics, there is a cycle of trajectories (i.e. heteroclinic cycle) connecting the pure strategy equilibria. In (a),  the cycle is repelling and the community persists. In (b),  the cycle is attracting and the community is extinction prone. Simulated metapopulations consist of $30$ patches with all-to-all coupling for dispersing individuals and spatial variation in payoffs ($c^r=1+(r-1)/30$, $b^r =0.85\, c^r$, $a=3$). The fraction dispersing equals $d=0.005$ in (a) and $d=0.5$ in (b). }\label{fig:hetero}
\end{center}
\end{figure}

% see files lottery-bif-slice.R for the code that created these figures. 

\subsection*{Local coexistence}
We begin by studying the behavior of the within-patch dynamics (\ref{local})  in the absence of dispersal. If only strategy $i$
is present in patch $r$, then the per-capita growth rate of the strategy, call it $j$, dominated by strategy $i$  is $-m\, c_i^r/a$. Alternatively, the per-capita growth rate of the strategy $j$ dominating strategy $i$ equals $m \,b_i^r/a$. The three single-strategy equilibria  are connected by population trajectories in which dominant strategies replace subordinate strategies (Fig.~\ref{fig:hetero}). This cycle of
population trajectories in patch $j$ is known as a \emph{heteroclinic
  cycle} \citep{hofbauer-sigmund-98}. Using average Lyapunov functions,
time-one maps, or measure-theoretic
techniques \citep{hofbauer-81,krupa-melbourne-95,jde-00}, one can show that the strategies in patch $r$ locally coexist in the sense of permanence provided that the product of the invasion rates exceeds the product of the exclusion rates:
\begin{equation}\label{local permanence}
\prod_{i}b_i^r> \prod_{i} c_i^r.
\end{equation}
Interestingly, inequality \eqref{local permanence} is equivalent to the determinant of the payoff matrix being positive.

When coexistence occurs, the heteroclinic cycle of the boundary of the population state space is repelling and there is a positive global attractor for the within-patch dynamics  (Fig.~\ref{fig:hetero}a) . When inequality \eqref{local permanence} is reversed, the heteroclinic cycle on the boundary is attracting (Fig.~\ref{fig:hetero}b). The strategies asymptotically cycle between three states (rock-dominated, paper-dominated, scissor-dominated), and the frequencies of the under-represented strategies asymptotically approach zero. Hence, all but one strategy goes extinct when accounting for finite population sizes.

\subsection*{Metacommunity coexistence.}

\noindent{ \bf Analytical results.}
When the patches are coupled by dispersal,  we show in Appendix A that for any pair of strategies, the dominant strategy competitively excludes the subordinate strategy. Hence, as in the case of the dynamics within a single patch, the metacommunity exhibits a heteroclinic cycle  on the boundary of the metacommunity phase space. 

Work of \citet{jde-10} on permanence for structured populations (see Appendix B) implies that metapopulation persistence is determined by invasion rates and exclusion rates at single strategy equilibria. More specifically, consider the rock strategy equilibrium where $x_1^r =1$ and $x_2^r=x_3^r=0$ for all patches $r$. Linearizing the paper strategy dynamics at the rock equilibrium yields 
\[
\frac{dx_2^r}{dt}\approx -m\, x_2^r +m \frac{\sum_s d_{sr} (a+ b_2^s) x_2^s}{\sum_ s d_{sr} a}. 
\]
Equivalently, if $\x_2 = (x_2^1,\dots, x_2^n)^T$ where $^T$ denotes transpose, then
\[
\frac{d\x_2}{dt} \approx \left(-m I+m \Psi D^T (aI+B_2) \right) \x_2
\]
where $I$ is the identity matrix, $\Psi$ is the diagonal matrix with entries $1/\sum_ s d_{1s} a^s,\dots,1/\sum_s d_{ns} a^s$,  $B_2$ is the diagonal matrix with diagonal entries $b_2^1,\dots,b_2^n$, and $D^T$ is the transpose of the dispersal matrix. Corresponding to the fact that the paper strategy can invade the rock strategy, the stability modulus of $-mI+m\Psi D^T(aI+B_2)$ (i.e. the largest real part of the eigenvalues) is positive. Call this stability modulus $\mathcal{I}_2$, the invasion rate of strategy $2$. Linearizing the scissor strategy dynamics at the rock equilibrium yields 
\[
\frac{d\x_3}{dt} \approx \left(-mI+m \Psi D^T (aI-C_3) \right) \x_3
\]
where $C_3$ is the diagonal matrix with diagonal entries $c_3^1,\dots,c_3^n$. Corresponding to the fact that scissor strategy is displaced by the rock strategy, the stability modulus of $-mI+m\Psi D^T (aI-C_3)$ is negative. We call this negative of this stability modulus ${\mathcal E}_3$, the exclusion rate of strategy $3$. By linearizing around the other pure strategy equilibria, we can define the invasion rates ${\mathcal I}_i$ for each strategy invading its subordinate strategy and the exclusion rates ${\mathcal E}_i$ for each strategy being excluded by its dominant strategy. 

Appendix A shows that the metapopulation persists if the product of the invasion rates exceeds the product of the exclusion rates:
\begin{equation}\label{eq:condition}
\prod_{i=1}^3 \mathcal{I}_i > \prod_{i=1}^3 \mathcal{E}_i
\end{equation}
If the inequality \eqref{eq:condition} is reversed, then the metapopulation is extinction prone as initial conditions near the boundary converge to the heteroclinic cycle and all but one strategy is lost regionally. While inequality~\eqref{eq:condition} can be easily evaluated numerically, one can not, in general, write down a more explicit expression for this permanence condition. However, when the metapopulation is weakly mixing (i.e. dispersal rates are low) or well-mixed (i.e. dispersal rates are high), we are able to find more explicit criteria. Furthermore, when dispersal is unconditional, we show that there is critical dispersal rate below which persistence is possible (Appendix C).

At sufficiently low dispersal rates i.e $d_{rr}\approx 0$ for all $r$,  the metacommunity coexistence criterion~\eqref{eq:condition} simplifies to
\begin{equation}\label{metacommunity permanence low}
\prod_{i=1}^3\max_r b_i^r
>
\prod_{i=1}^3 \min_r c_i^r.
\end{equation}
Unlike the local coexistence criterion \eqref{local permanence} which requires  that the geometric mean of benefits exceeds the geometric mean of costs within a patch, inequality \eqref{metacommunity permanence low} requires that the geometric mean of the maximal benefits exceeds the geometric mean of the minimal costs. Here, the maxima and minima are taken over space.  Thus, inequality \eqref{metacommunity permanence low} implies that localized dispersal promotes  coexistence if there is sufficient spatial variation in relative benefits, costs, or mortality rates.

For well-mixed metacommunities (i.e.  $d_{rs}\approx v_s$ for all $r,s$), the invasion rate $\mathcal{I}_i$ of the strategy is  approximately $m \, \sum_r b_i^r/a$. Conversely, the exclusion rate $\mathcal{E}_i$ of strategy $i$ is  $-m\,\sum_r c_i^r/a$. These well-mixed metacommunities coexist provided that  the geometric mean of the spatially averaged benefit exceeds the geometric mean of the spatially averaged cost:
\begin{equation}\label{metacommunity permanence high}
\prod_{i=1}^3 \left(\frac{1}{n} \sum_r b_i^r \right)> \prod_{i=1}^3 \left(\frac{1}{n} \sum_r c_i^r\right).
\end{equation}
Since \eqref{metacommunity permanence high} implies \eqref{metacommunity permanence low}, it follows that persistence of well-mixed communities implies persistence of weakly-mixing communities, but not vice-versa. We can refine this observation under the assumption of unconditional dispersal. 

 Unconditional dispersal occurs when the fraction of individuals dispersing, $d$, is independent of location. Let $p_{rs}$ denote the fraction of dispersing individuals from patch $r$ that end up in patch $s$ i.e. $p_{sr}$ is a dispersal kernel that describes how dispersing individuals redistribute across patches. Under these assumptions, the fraction $d_{rs}$ of individuals in patch $r$ dispersing to patch $s\neq r$ equals $d\, p_{rs}$. The fraction $d_{rr}$ of individuals remaining in patch $r$ is $1-d$.  In Appendix C, we show that there is a critical dispersal threshold $d^*$ (possibly $0$ or $1$) such that the metacommunity persists if its dispersal rate is below $d^*$ and is extinction prone when its dispersal rate is greater than $d^*$. It follows that if the metacommunity persists when highly dispersive (i.e. $d^*=1$), then it  persists at all positive dispersal rates. Conversely,  if a metacommunity is extinction prone when weakly mixing (i.e. \eqref{metacommunity permanence low} is violated), then it is extinction prone at all positive dispersal rates. 

\noindent{ \bf Numerical results.}
To illustrate the implications our analytical results, we consider two scenarios where either there is only spatial variation in the payoffs or where there is within-patch and spatial variation in payoffs. There are $n=30$ patches  that are equally connected. A fraction $d$ of individuals disperse and dispersing individuals are distributed equally amongst the remaining patches (i.e. $d_{rs}=d/(n-1)$ for $r\neq s$). For this form of dispersal, the metapopulation is well-mixed when $d=(n-1)/n$ in which case $d_{rs}=1/n$ for all $r,s$. 

First, we consider the case where there is spatial variation in payoffs, but  all strategies within a patch fare equally well when they are the dominant player in an interaction and fare equally poorly when they are the subordinate player in the interaction (i.e.  $b_i^r =b^r$, and $c_i^r=c^r$ for all $i=1,2,3$). Local coexistence requires that the benefit $b^r$ to the winner must exceed the cost $c^r$ to the loser. For well-mixed communities, regional coexistence requires that
the spatially averaged benefit $\frac{1}{n}\sum_r b^r$ must exceed the spatially averaged average cost $\frac{1}{n} \sum_r c^r$.  From these two conditions, it follows that metapopulation persistence for well-mixed communities requires that at least one of the patches promotes local coexistence.

\begin{figure}[t]
\begin{center}
\begin{tabular}{cc}
\includegraphics[height=3.5in,trim=1cm 1cm 2cm 1cm,clip]{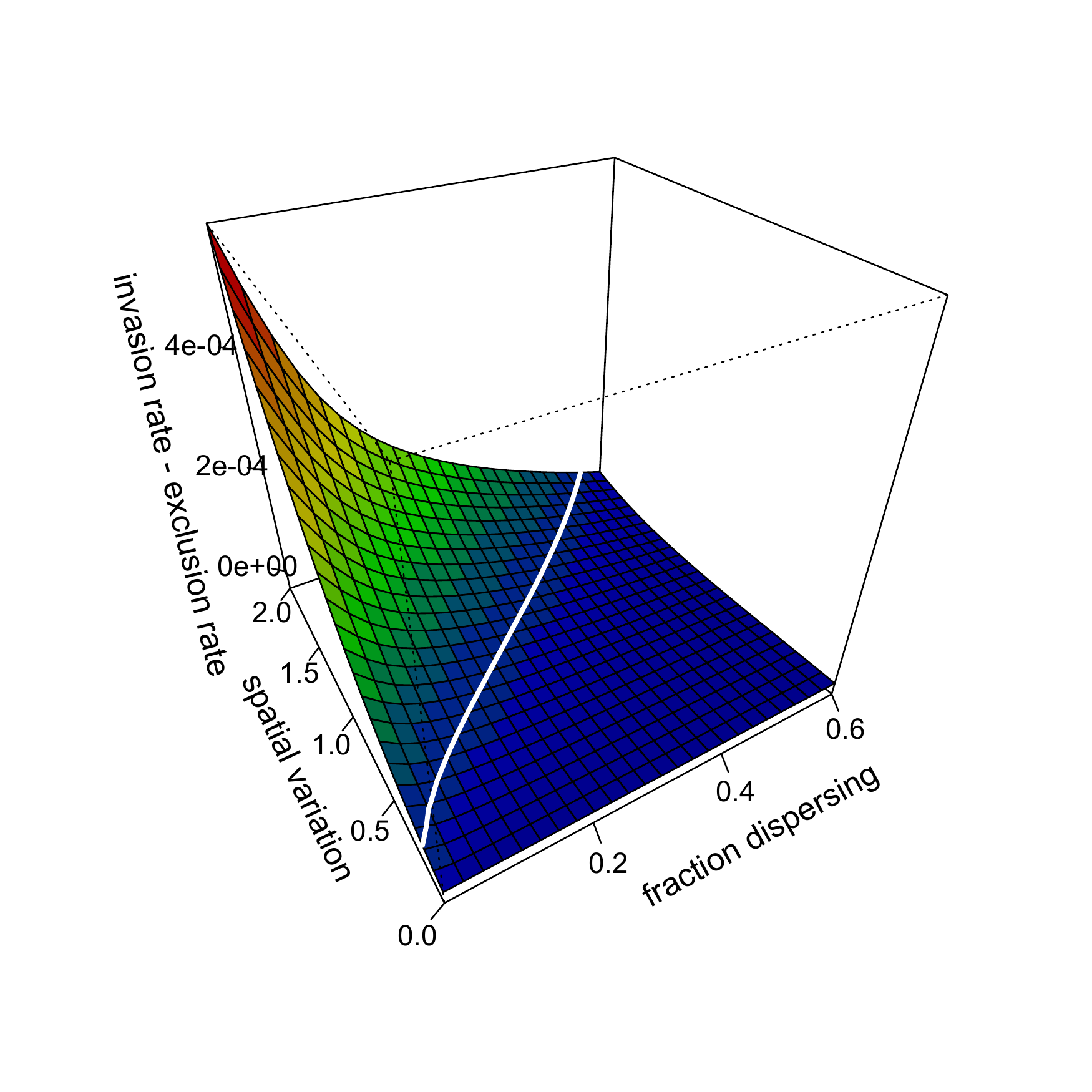}
&\includegraphics[height=3.5in,trim=1cm 1cm 2cm 1cm,clip]{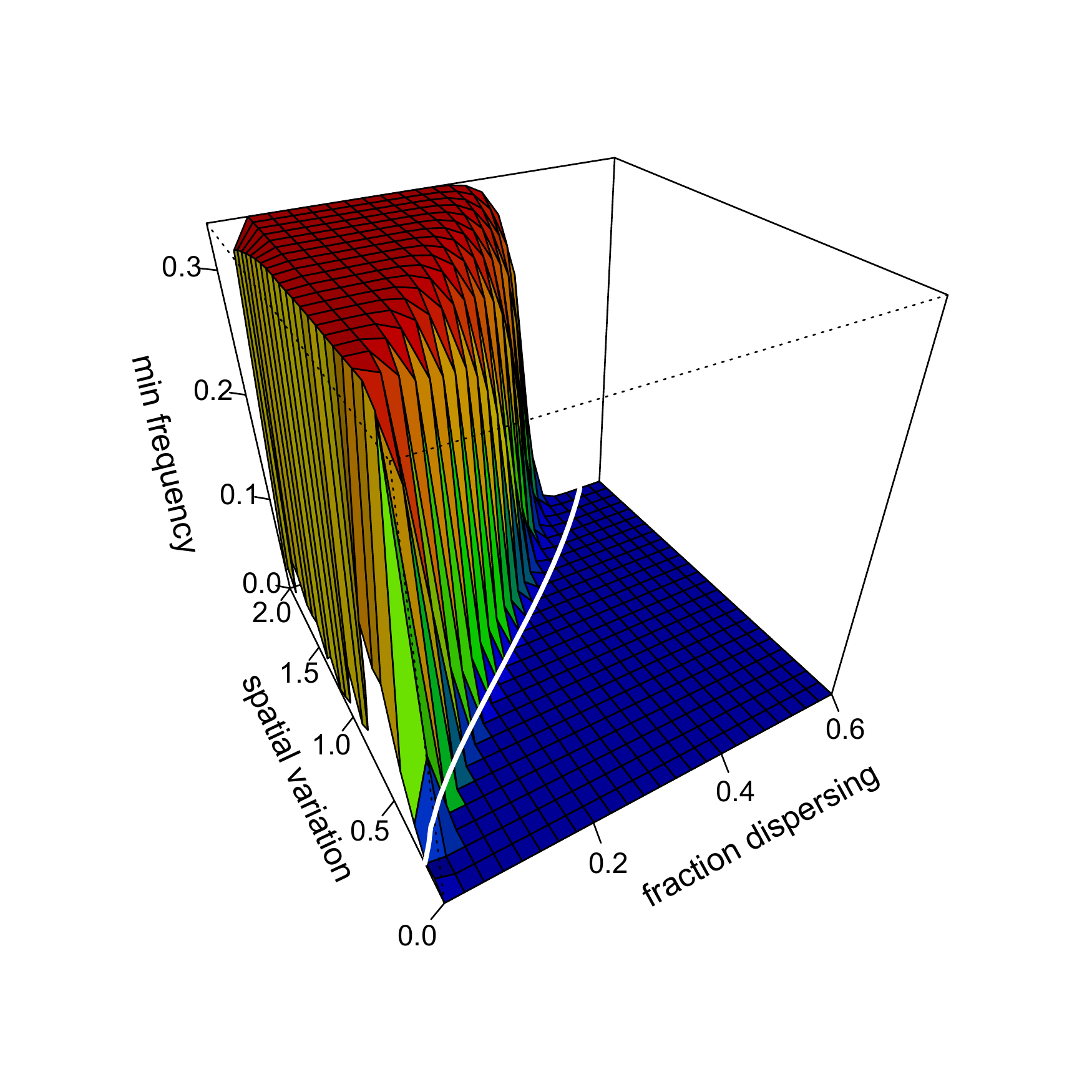}\\
(a) &(b) \\
\end{tabular}
\caption{The effect of spatial variation and dispersal rate on the persistence criterion in (a) and  the long-term metapopulation frequencies in (b). Metapopulations consist of $30$ patches with all-to-all coupling for dispersing individuals and spatial variation in payoffs ($c^r=1+(r-1)\sigma/30$, $b^r =0.85\, c^r$, $a=3$).  In (a), the difference between the product $\prod_i \mathcal{I}_i$ of the invasion rates and the product $\prod_i \mathcal{E}_i$ of the exclusion rates  are plotted as function of the fraction $d$ of dispersing individuals and the range $\sigma$ of spatial variation in the payoffs. Positive values correspond to persistence and negative values to the metapopulation being extinction prone. The white curve is where the difference in products equals zero. In (b), the minimal and maximal frequencies for one patch and the spatial average  are plotted as a function of the fraction $d$ of dispersing individuals and  $\sigma$. The white curve is where the difference in the products of invasion and exclusion rates  equals zero.  }\label{fig:bif2D}
\end{center}\end{figure}

\begin{figure}[t]
\begin{center}
\includegraphics[width=6.5in]{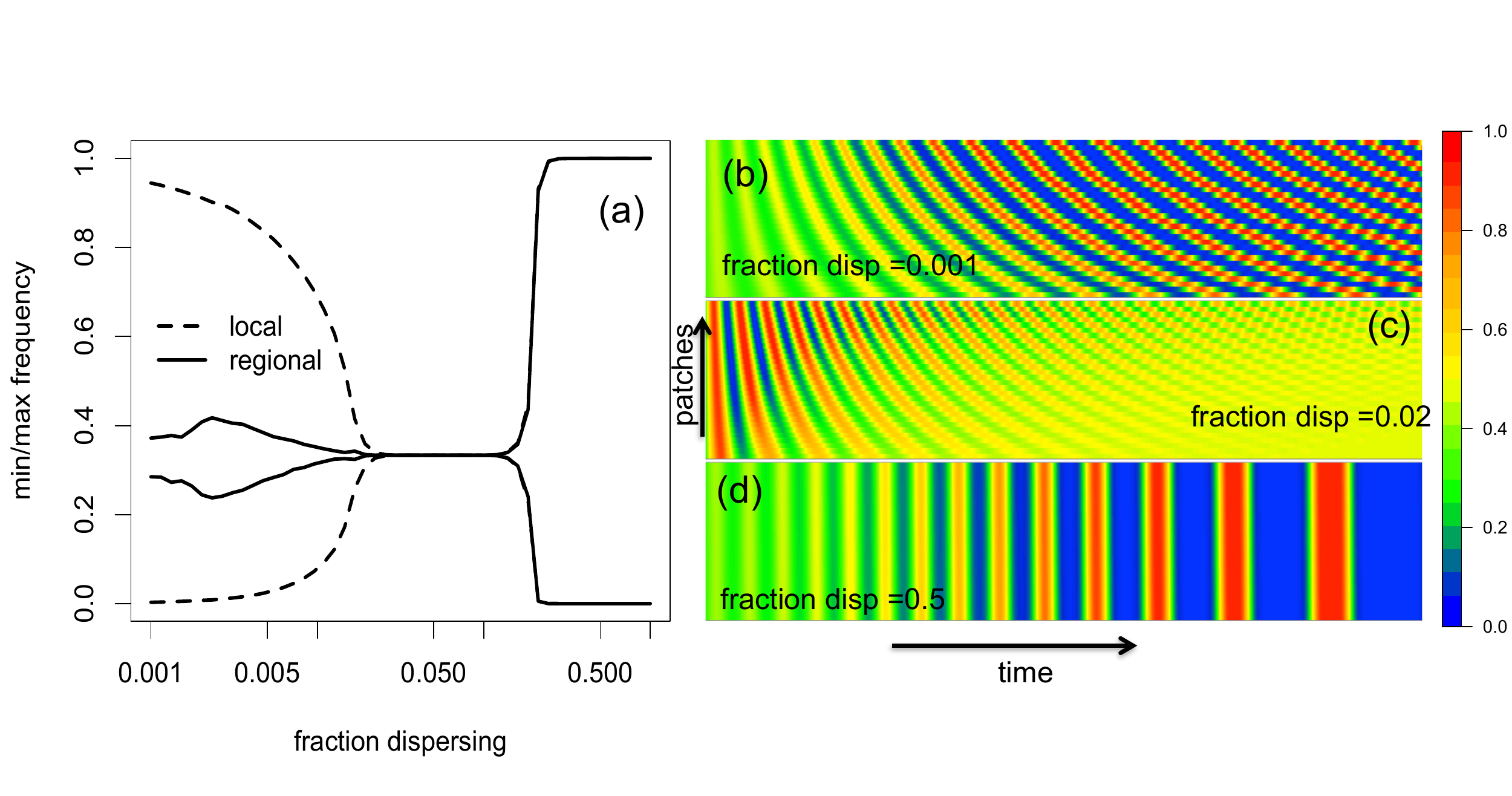}
\caption{The effect of dispersal rates on metapopulation dynamics. Metapopulations consist of $30$ patches with all-to-all coupling for dispersing individuals and spatial variation in payoffs ($c^r=1+(r-1)/30$, $b^r =0.85\, c^r$, $a=3$).  In (a), the minimal and maximal frequencies for one patch and the spatial average  are plotted as a function of the fraction $d$ of dispersing individuals. In (b)-(d), the spatial-temporal dynamics are plotted for low, intermediate, and high dispersal rates. Rock frequencies are color-coded as indicated.  }\label{fig:bif1D}
\end{center}\end{figure}

When all patches fail to promote local coexistence (i.e. $c^r>b^r$ for all $r$), weakly mixing metacommunities persist provided that the benefit in some patch exceeds the cost in another (possibly the same) patch i.e.
$
 \max_r b^r
>
 \min_r c^r.
$
When this condition is meet, there is a critical dispersal threshold $d^*$ below which the metacommunity persists, and above which the metacommunity is extinction-prone.  

Figure~\ref{fig:bif2D}a demonstrates the analytical prediction that the difference between the products of the invasion and exclusion rates is a decreasing function of the fraction $d$ dispersing. Furthermore, the difference in products is an increasing function of the amplitude of the spatial variation in payoffs.  Hence, the critical dispersal threshold  increases with the amplitude of the spatial variation of the payoffs. Intuitively, higher dispersal rates are needed to average out greater spatial variation. Unlike the difference between the products of invasion and exclusion rates, the minimum frequency of strategies exhibits a highly nonlinear response to increasing dispersal rates (Fig~\ref{fig:bif2D}b): the minimal frequency initially increases with dispersal rates, reaches a plateau of approximately one-third at intermediate dispersal rates, and decreasing abruptly to zero after crossing the critical dispersal. 

At low dispersal rates, metacommunity persistence is achieved  by a spatial game of hide and seek  (Figs.~\ref{fig:bif1D}a,b). At any point in time, each strategy is at high frequency in some patches and low frequencies in the remaining patches. Strategy composition in each patch cycles as dominant strategies displace subordinate strategies. Intermediate dispersal rates stabilize the local and regional dynamics (Figs.~\ref{fig:bif1D}a,c). As a consequence, local diversity is maximal at intermediate dispersal rates. At high dispersal rates, the population dynamics synchronize across space as they approach the heteroclinic cycle (Figs.~\ref{fig:bif1D}a,d). 

\begin{figure}[t]
\begin{center}
\begin{tabular}{cc}
\includegraphics[height=3.5in,trim=1cm 1cm 2cm 1cm,clip]{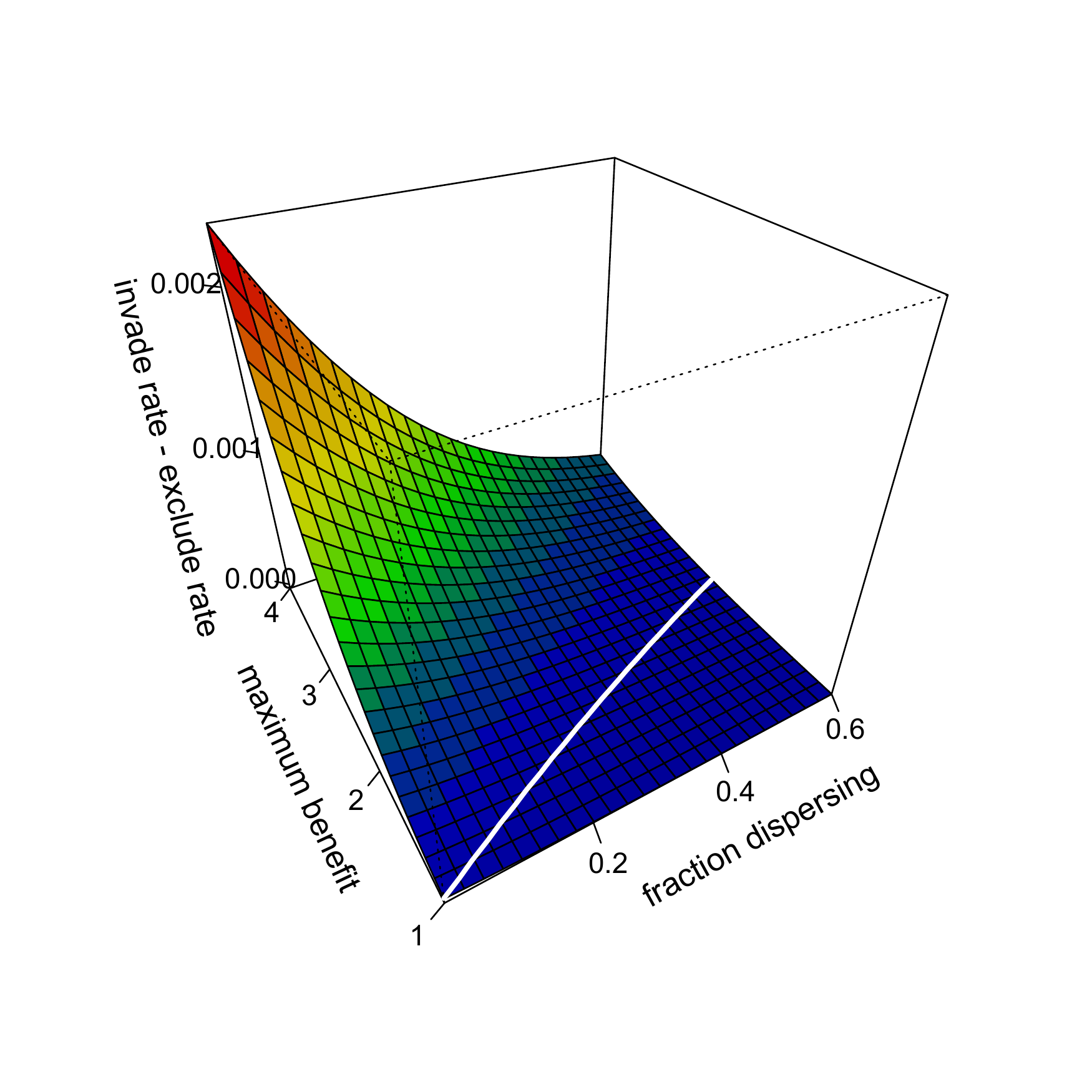}&\hskip-0.1in\includegraphics[height=3.5in,trim=1cm 1cm 2cm 1cm,clip]{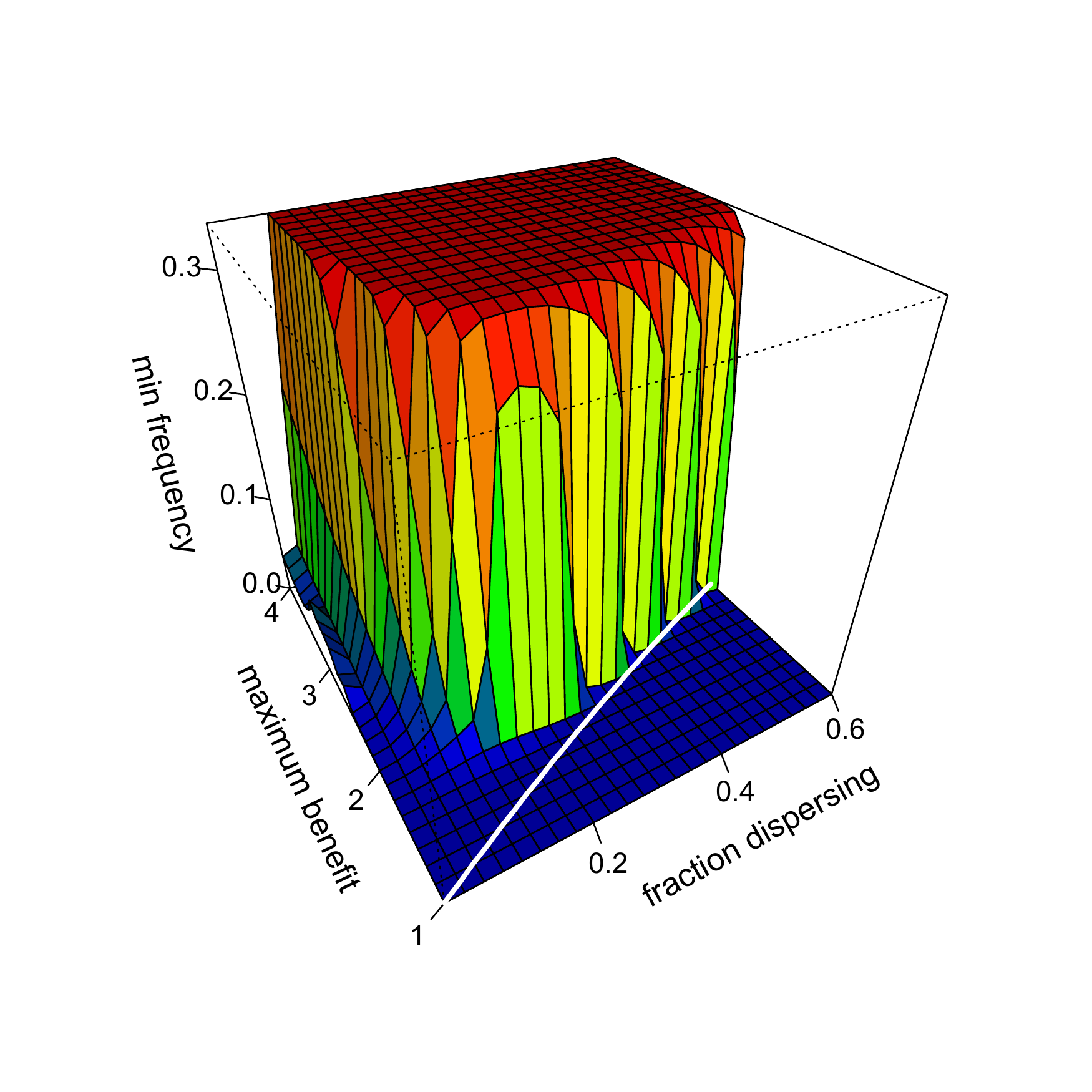}\\
(a) &(b) \\
\end{tabular}
\caption{The effect of spatial variation and dispersal rate on the persistence criterion in (a) and long-term metapopulation frequencies in (b). Metapopulations consist of $30$ patches with all-to-all coupling for dispersing individuals.  Each strategy has $10$ patches in which their benefit equals $b_{high}$ and equals $0$ in the remaining patches. $c=1$, $a=2$, $m=0.1$ in all patches.  In (a), the difference between the product $\prod_i \mathcal{I}_i$ of the invasion rates and the product $\prod_i \mathcal{E}_i$ of the exclusion rates  are plotted as function of the fraction $d$ of dispersing individuals and the maximal benefit $b_{high}$. Positive values correspond to persistence and negative values to the metapopulation being extinction prone. The white curve is where the difference of products equals zero. In (b), the minimal and maximal frequencies for one patch and the spatial average  are plotted as a function of the fraction $d$ of dispersing individuals and the maximal benefit  $b_{high}$. The white curve is where the difference in the products of invasion and exclusion rates equals zero. }\label{fig:bif2D-2}
\end{center}\end{figure}

\begin{figure}[t]
\begin{center}
\includegraphics[width=6.5in]{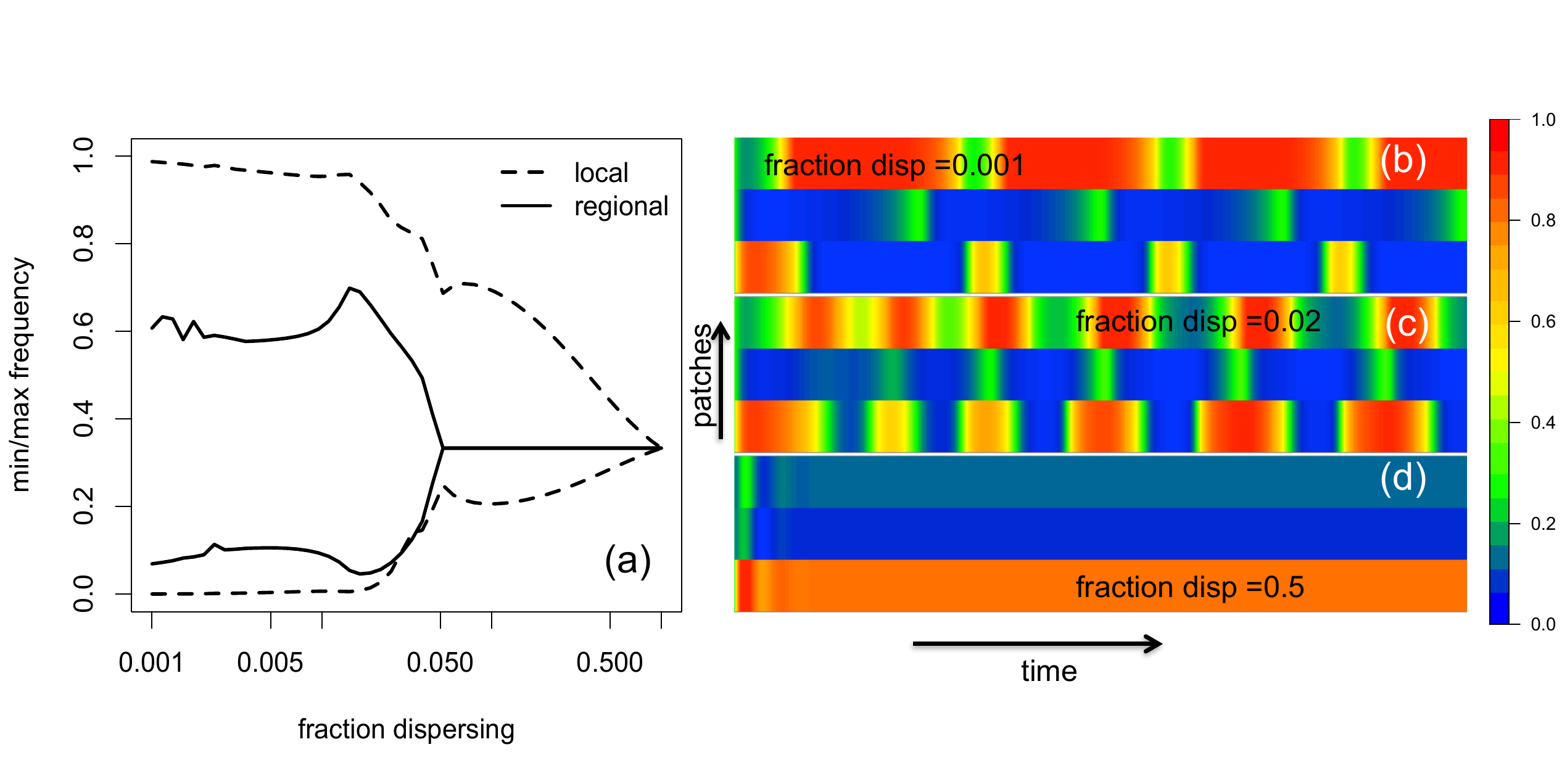}
\caption{The effect of dispersal rates on metapopulation dynamics. Metapopulations consist of $30$ patches with all-to-all coupling for dispersing individuals. Each strategy has $10$ patches in which their benefit equals $b_{high}=4$ and equals $0$ in the remaining patches. $c=1$, $a=2$, $m=0.1$ in all patches.  In (a), the minimal and maximal frequencies for one patch and the spatial average  are plotted as a function of the fraction $d$ of dispersing individuals. In (b)-(d), the spatial-temporal dynamics are plotted for low, intermediate, and high dispersal rates. Rock frequencies are color-coded as indicated.}\label{fig:bif1D-2}
\end{center}\end{figure}

For the second numerical scenario, we consider when payoffs vary within patches  (e.g. rock gets a higher benefit than scissor when playing their subordinate opponents in one patch, but scissor gets the higher benefit in another patch) as well as spatially. In this case, well-mixed communities can persist despite being locally extinction prone.  To understand why, assume each strategy wins big in some patches but win nothing in others. Let $f$ denote the fraction of patches where a strategy wins big and receives a payoff $b_{high}$ against its subordinate strategy.  In the remaining fraction $1-f$ of patches, each strategy receives no benefit  when playing against their subordinate strategy. Furthermore,  assume that there is no variation in the costs $c_i^r=c$ for all $i,r$. Under these assumptions, local coexistence is impossible as $c>b_{high}\cdot 0 = 0$. In contrast, a well-mixed metacommunity persists if 
$ f\, b_{high} > c$ and a weakly-mixing metacommunity persists if $b_{high}>c$.  Therefore, provided that $b_{high}$ is sufficiently large,  coupling the communities by any level of dispersal mediates regional coexistence despite local communities being extinction prone.

Consistent with these analytical predictions, Fig.~\ref{fig:bif2D-2} illustrates that metacommunity persists at all dispersal rates if the difference in payoffs  is sufficiently great ($b_{high}>3$) and only persists at low dispersal rates for intermediate differences in the payoffs. When there are large differences, metapopulation abundance and stability increases continually with dispersal rates (Fig.~\ref{fig:bif1D-2}). In contrast, metapopulation abundance is maximized at intermediate dispersal rates whenever there are intermediate differences in the payoffs (Fig.~\ref{fig:bif2D-2}b).
 
\section*{Discussion} 

The rock-paper-scissors game represents the prototypical situation in which the components of a system satisfy a set of non-transitive relations. It is a surprising and fascinating feature of recent work in evolutionary biology and ecology that such interactions have been discovered in a wide range of natural systems~\citep{buss-jackson-79,sinervo-lively-96,kerr-etal-02,kirkup-riley-04,lankau-strauss-07,cameron-etal-09}. The existence of non-transitive interactions in biological systems has been suggested as an important mechanism for maintaining biodiversity~\citep{durrett-levin-97,kerr-etal-02,lankau-strauss-07,roelke-eldridge-10,allesina-levine-11}. This suggestion, however, raises an important theoretical question: Is it possible for all components of such a system to persist in the long term? This question is pertinent since modeling the dynamics of the rock-paper-scissors game (and related non-transitive systems) using the replicator equation shows that cyclic behavior corresponds to convergence toward a heteroclinic attractor on the boundary of the strategy space, and this process must ultimately result in the extinction of some strategies~\citep{hofbauer-sigmund-98}.

It is widely believed in ecology that the inclusion of spatial structure, in which the interactions of individuals are local, can result in the coexistence of communities that could not persist in a panmictic situation~\citep{durrett-levin-97,hanski-99,amarasekare-nisbet-01,holyoak-etal-05}. There are numerous ways in which a spatially structured population can be modeled mathematically, depending on the assumptions made regarding the nature of the spatial interactions of the individuals in the population~\citep{durrett-levin-94}. Possible approaches include reaction-diffusion systems~\citep{cantrell-cosner-03}, metapopulation and metacommunity theory~\citep{hanski-99,holyoak-etal-05}, coupled lattice maps~\citep{hastings-93,holland-hastings-08}, and cellular automata and related lattice models~\citep{nowak-may-92,killingback-doebeli-96,durrett-levin-97,durrett-levin-98,iwasa-etal-98,kerr-etal-02}.   

Most previous attempts to understand the effect of spatial structure on the persistence of systems with non-transitive interactions have utilized cellular automata-type models~\citep{durrett-levin-97,durrett-levin-98,iwasa-etal-98,frean-abraham-01,kerr-etal-02,karolyi-etal-05,reichenbach-etal-07,rojas-allesina-11}.  The main conclusion that can be drawn from these cellular automata studies is that in three-species systems with non-transitive interactions it is possible for all species to coexist in a spatially structured model even when they could not all persist in the corresponding panmictic system. Coexistence in these models when formulated in two spatial dimensions results from the different species aggregating in regions that cyclically invade each other.  It is worth noting that in the reaction-diffusion approach of \citet{nakamaru-iwasa-00} coexistence is not possible in one-dimensional systems. This issue has not, however, been investigated using lattice models. Cellular automata models have the virtue of explicitly introducing space through a lattice of cells and of directly modeling the spatial interactions between individuals. However, such models also have a number of significant limitations. Since spatial structure is introduced in a very concrete fashion, through an explicit choice of a spatial lattice (almost always taken to be a two-dimensional square lattice) and a spatial interaction neighborhood (usually taken to be the eight cells surrounding the focal cell) it is, in general, unclear how changes in these structures affect species coexistence. A second limitation of cellular automata models is the difficulty is using them to study the effects of spatial heterogeneity. In all the lattice models of non-transitive interactions that have been studied the rules determining how cells are updated are the same at every spatial location, although it is known, in general, that spatial heterogeneity may have important implications for species coexistence~\citep{amarasekare-nisbet-01}. A third limitation is that cellular automata are notoriously difficult to study analytically, and indeed almost all the key results on coexistence of species with non-transitive interactions in lattice models have been obtained from simulations (see, however, \citet{durrett-09}).   

In this paper we have adopted the metacommunity perspective to formulate a new approach to studying the dynamics of spatially structured communities in which rock-paper-scissors-type interactions hold. This approach assumes that the overall metacommunity is composed of a number of local communities, within each of which the interactions are panmictic, and that the local populations are coupled by dispersal. The resulting metacommunity model allows for a very general treatment of the population dynamics of spatially structured systems with non-transitive interactions, which overcomes many of the limitations inherent in cellular automata-type models. In particular, our model allows a very general treatment of dispersal between spatial patches, includes spatial heterogeneity in a fundamental way, and allows precise analytic derivations of the central results.   

In our model, in the absence of dispersal, the population dynamics within each patch exhibits
a heteroclinic cycle.  Coexistence of all strategies in any given patch requires that the geometric
mean of the benefits obtained from the payoff exceed the geometric mean of the costs within that
patch. Moreover, when the spatial patches are coupled by dispersal the metacommunity possesses a
heteroclinic cycle, and all members of the metacommunity persist when a regional coexistence criterion holds--the geometric mean of invasion rates when rare of the dominant strategies exceed the geometric mean of the exclusion rates when rare of the subordinate strategies. Although it is not possible, in general, to write down an explicit formula for the eigenvalues associated with these invasion and exclusion rates, it is possible to find more explicit expressions in the limiting cases of weakly-mixed metacommunities and well-mixed metacommunities. Weak mixing occurs when when dispersal rates are low.  In this case, our analysis reveals that sufficient spatial heterogeneity in the payoffs for pairwise interactions allows metacommunity coexistence even when every local community is extinction prone. Thus, in the presence of spatial heterogeneity, local dispersal promotes coexistence. Alternatively, when dispersal rates are high,  the metacommunity is well-mixed. In this case, the coexistence criterion requires that the geometric mean of spatially averaged benefits obtained from the payoff exceed the geometric mean of the spatially averaged costs. These coexistence criteria  imply that the coexistence of a well-mixed metacommunity guarantees the coexistence of the corresponding weakly mixed one.   The converse result does not hold.   Thus, metacommunities with higher dispersal rates are less likely to persist than those with lower ones.

For unconditional dispersal (i.e. when the fraction $d$ of individuals dispersing is independent of location), the interaction between spatial heterogeneity and dispersal leads to a threshold effect: there exists a critical dispersal value $d^*$, such that if the dispersal rate is less than $d^*$ the metacommunity persists, while if the dispersal rate is greater than $d^*$ it is extinction prone. This threshold effect occurs whenever well-mixed communities are extinction prone but weakly-mixed communities are not. For example, there is sufficient spatial variation in the payoffs but the cost paid by the loser exceeds the benefit gained by the winner in every pairwise interaction. Similar dispersal thresholds have been demonstrated for two-species competitive communities exhibiting either priority effects or local competitive dominance~\citep{levin-74,amarasekare-nisbet-01}. However, unlike these transitive systems, regional coexistence for these intransitive systems does not require each species having regions in space where either they are initially more abundant or competitively dominant.  

Our results on the effect of dispersal on the coexistence of rock-paper-scissors metacommunities are in broad qualitative agreement with the conclusions that can be drawn from cellular automata-type models that include the movement of individuals, which is the lattice analogue of dispersal. \citet{karolyi-etal-05} considered a two-dimensional lattice model of non-transitive interactions in which individuals moved due to a chaotic flow, such as might occur in a fluid system. \citet{reichenbach-etal-07} also studied the effect of mobility on coexistence in a two-dimensional cellular automata model of rock-paper-scissors interactions, where individual movement was modeled using techniques of dimmer automata~\citep{schofisch-hadeler-96}. In each case it was found through simulation that there exists a critical level of mobility, below which all species coexist and above which only one species survives in the long-term. This critical mobility level in lattice models of rock-paper-scissors interactions is the analogue of the critical dispersal rate $d^*$ in our metacommunity model. It is interesting to note in this context that a similar threshold also occurs in a model of cyclic interactions on complex networks studied by \citet{szabo-etal-04}. In this case if the fraction of long-range interactions present in a small-world network is below a critical value coexistence of all species is possible, while if it is exceeded species extinctions occur.

We also note that a further example of a lattice model that has been used to study the effect of spatial structure in maintaining meta-community persistence in a system with non-transitive interactions occurs in the area of prebiotic evolution. \citet{eigen-schuster-79} observed that there is a fundamental problem in the evolution of self-replicating molecules: there exists an information threshold since the length of the molecule is restricted by the accuracy of the replication process. Eigen and Schuster proposed as a solution to this problem the concept of the hypercycle, in which a number of molecules catalyze the replication of each other in a cyclic fashion. The dynamics of a hypercycle can be modeled mathematically as a replicator equation with a cyclic payoff matrix~\citep{hofbauer-sigmund-98}, and thus the hypercycle corresponds dynamically to a replicator system with non-transitive interactions. The concept of a hypercycle has, however, a crucial flaw: it is not evolutionarily stable - selection will favor the evolution of a parasitic mutant which does not provide any catalytic support to other molecules in the hypercycle even though it receives such catalytic support itself~\citep{maynard-smith-79,bresch-etal-80}. The evolution of parasitic mutants results in the collapse of hypercycles as entities capable of encoding information. Interestingly, the inclusion of spatial structure can prevent the evolution of selfish mutants and may result in the persistence of hypercycles. The effect of spatial structure on the persistence of hypercycles has been studied using a cellular automaton model in~\citep{boerlijst-hogeweg-91}. It is shown in this model that local spatial interactions result in the formation of self-organized spiral waves, and that selection acting between these spiral waves can counteract the effect of selection acting at the level of the individual molecules, with the consequence that the hypercycle can be resistant to the evolution of parasitic mutants.

The metacommunity model we have introduced here provides a complementary approach to the lattice models that have previously been used to study coexistence in rock-paper-scissors-type systems. It seems likely that each type of model will most naturally describe different types of empirical systems with non-transitive interactions. For example, the lattice modeling approach may describe reasonably well an \textit{in vitro} microbial population growing on a plate~\citep{kerr-etal-02}. In contrast, our metacommunity model would seem to be a more natural approach to use to describe an \textit{in vivo} microbial population inhabiting many host organisms with transmission between the hosts, as in the model system of \citet{kirkup-riley-04}, or plant communities living on different soil types~\citep{lankau-strauss-07,cameron-etal-09}. This observation raises the possibility that it may be possible to use such systems to empirically test the predictions of our metacommunity model.

\bibliography{../../seb}
\newpage

\include{Cyclinginspace9Appendix}

\end{document}